\documentclass[a4paper,11pt]{article}
\pdfoutput=1 

\usepackage{jinstpub} 
\usepackage{caption}
\usepackage{subcaption}

\title{High precision scalable power converter for accelerator magnets}

\author[a,b,1]{Krister Leonart Haugen,\note{Corresponding author.}}
\author[b]{Konstantinos Papastergiou,}
\author[b]{Panagiotis Asimakopoulos}
\author[a]{Dimosthenis Peftitsis}

\affiliation[a]{Department of Electric Power Engineering, Norwegian University of Science and Technology (NTNU),\\Trondheim, Norway}
\affiliation[b]{Medium Power Converters Section, CERN,\\Geneva, Switzerland}

\emailAdd{krister.leonart.haugen@cern.ch}

\abstract{The lower conduction power losses and the positive temperature coefficient that favours parallel connections, make Silicon Carbide (SiC) metal oxide semiconductor field-effect transistors (MOSFETs) to be an excellent replacement of existing Silicon insulated gate bipolar transistors (IGBTs) technology. These characteristics combined with high switching frequency operation, enables the design of high-accuracy DC-DC converters with minimised filtering requirements. This paper investigates the design for a converter with high-accuracy current (0.9ppm) supplying a 0.05H electromagnetic load, aiming to achieve the accuracy without the use of active filters, by using SiC MOSFETs and a scalable module-based converter design.}

\arxivnumber{2201.10824} 

\proceeding{TWEPP 2021 Topical Workshop on Electronics for Particle Physics\\
  20 to 24 September 2021\\
  Online}
\begin{document} 
\maketitle
\flushbottom

\section{Introduction}
\label{sec:intro}
A primary objective of the powering in particle accelerator is the precision and reproducibility of the experiments. Scalable converters enable sub-ppm current precision with minimal filtering requirements by using multiple modules in interleaved operation \cite{Gorji2019}. In order to reach the <1ppm level which is required for some applications, active filters are often used \cite{Parchomiuk2016}. However, they require relatively large reactors, contribute to power losses and their control is not trivial. With the introduction of  Silicon Carbide (SiC) metal oxide semiconductor field-effect transistors (MOSFETs) as a commercial alternative to silicon insulated gate bipolar transistors (IGBTs) for switch-mode converters, it is now feasible to utilise SiC semiconductor technology in the world of high-energy particle physics as well. This paper demonstrates a case study for a 0.05H electromagnet with high accuracy requirement of 0.9ppm and investigates the performance improvements to the design and operation of such converters when using SiC MOSFETs compared to IGBTs. The benefits of the SiC MOSFET are two-fold. Firstly, the reduced conduction on-state losses and positive on-state temperature coefficient and secondly, the reduced switching losses, enabling operation at higher switching frequencies. And since the SiC based technology is emerging with lower current ratings than the IGBTs, parallel connection is necessary to take advantage of the SiC MOSFETS.

The reduced on-state losses of SiC MOSFETs enables high-efficiency operation. Besides, their positive temperature coefficient facilitates a robust and more reliable parallel connection, as the positive temperature coefficient leads to natural balance of the current. If the parallelisation required to handle the current is achieved by parallel connecting H-bridges, introducing a number of parallel connected DC-DC converter enables the use of interleaving to achieve a lower current ripple.

\begin{figure}[!htp]
    \begin{subfigure}[t]{0.49\linewidth}
        \centerline{\includegraphics[width=\linewidth]{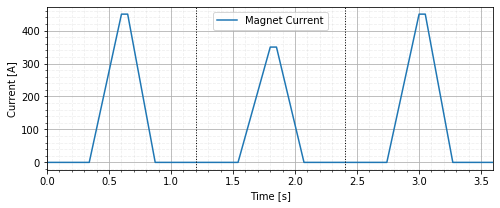}}
        \caption{Typical current cycle sequence, dotted lines indicate division between cycles.}
        \label{fig:current_mag}
    \end{subfigure}
    \begin{subfigure}[t]{0.5\linewidth}
        \centerline{\includegraphics[width=\linewidth]{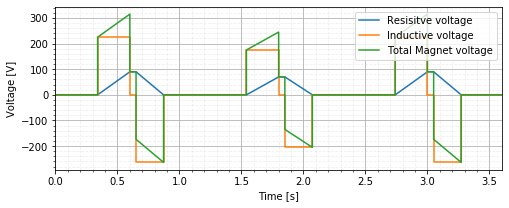}}
        \caption{Typical voltage requirement for the cycles.}
        \label{fig:voltage_mag}
    \end{subfigure}
    \caption{Examples of load which the DC-DC converters have to be designed for.}
\end{figure}

The current these power converters are designed to supply, will often be a series of trapezoid current pulses. These are supplied by power converters with built in energy recovery to reduce the overall energy consumption of the experiments \cite{LamailleSTUDYRENOVATION,Asimakopoulos2019}. An example of such a current train is shown in Fig. \ref{fig:current_mag}. Since the ramping and energy recovery often will require negative voltage to be applied across the magnet during ramp-down, such as in Fig. \ref{fig:voltage_mag}, the power converters are required to have 4-quadrant operation and a energy storage facility. Taking into account the advantages of SiC based MOSFETs, and taking the need for parallelisation as an opportunity, some interesting possibilities emerge with regards to reducing the ripple of the current supplied to magnets.


\section{The modular power converter}

Considering the constraints mentioned in the introduction, the magnets are normally supplied by an H-bridge with an electrolytic capacitor bank as energy storage \cite{Asimakopoulos2019}. The converter is such that often one converter is sufficient, perhaps combining 2 or 4 in series, parallel or series-parallel combination to achieve the required outputs, depending on the specific magnet. By opting for a parallel connection of power converters, the size and power ratings of each power converter can be reduced. The converter is now consisting of smaller identical modules, which will be named bricks, and for this paper the brick size has been chosen as 100V and 100A. This choice is based on previous work relating to the sizing system and operating cost of such modular power converters \cite{paper1}. 

\begin{figure}[htbp]
\centering 
\includegraphics[width=\textwidth]{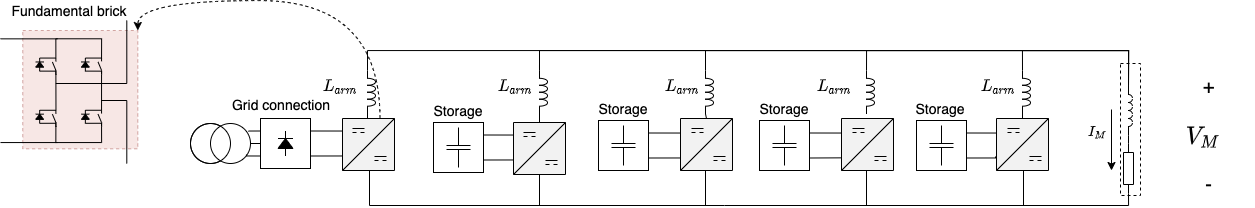}
\caption{\label{fig:topology} A possible configuration of the parallel connected bricks.}
\end{figure}

The proposed topology employs several bricks and assumes different functionality for them; some of them are connected to energy storage for the recovery of the magnet B-field energy while other bricks are connected to a power source (i.e. the grid) and supply the power losses due to the magnet resistance and the circuit self-consumption and the option to reduce the front-end peak-load and greater utilisation of the storage \cite{paper2}. More specifically in the example of Fig. \ref{fig:topology},  four of the bricks are connected to storage while just one brick is connected to the power grid via a 3-phase diode rectifier.

This paper ends up with the topology shown in Fig. \ref{fig:topology}, with each brick able to supply 100V and 100A. Four of the bricks are connected to storage, and supply the reactive power to the magnet, and the 5th brick is connected to the  grid via a 3-phase diode rectifier. Normally the power converter would be connected to the magnet via an active filter, which ensures the requirement for current accuracy. However, this filter is omitted in this paper in an effort to achieve the same accuracy by simply interleaving several bricks and operate them at higher switching frequencies. In this converter configuration and also in the modelling and simulations the arm inductance and the intrinsic inductances were considered small compared to the inductance of the load, meaning they don't need to be compensated for by the converter control. 

\section{Simulating the interleaving}

To investigate the performance of the parallel connected bricks, a PLECS model based of the topology in Fig. \ref{fig:topology} was developed and simulated, where the number of parallel connected bricks was varied from 1 up to 25, to investigate the effect of using SiC MOSFETs with relatively small current rating to supply a large current with interleaving. The single brick compares to the conventional case where the current is supplied by a single power converter. This was done to understand how the interleaving would influence the current ripple to the inductive load of a magnet. All simulations were performed without considering power losses, apart from the resistance in the magnet, or relevant limits in dv/dt for the power semiconductor devices and the magnets. The interleaving was achieved by phase-shifting each of the bricks evenly, i.e., the phase-shift of brick $N_i$ is $N_i = 360 \frac{1}{N_{bricks}}$. Where $N_{bricks}$ is the total number of bricks, and $N_i$ is one of the bricks.

\begin{figure}[htbp]
\centering 
\includegraphics[width=.5\textwidth]{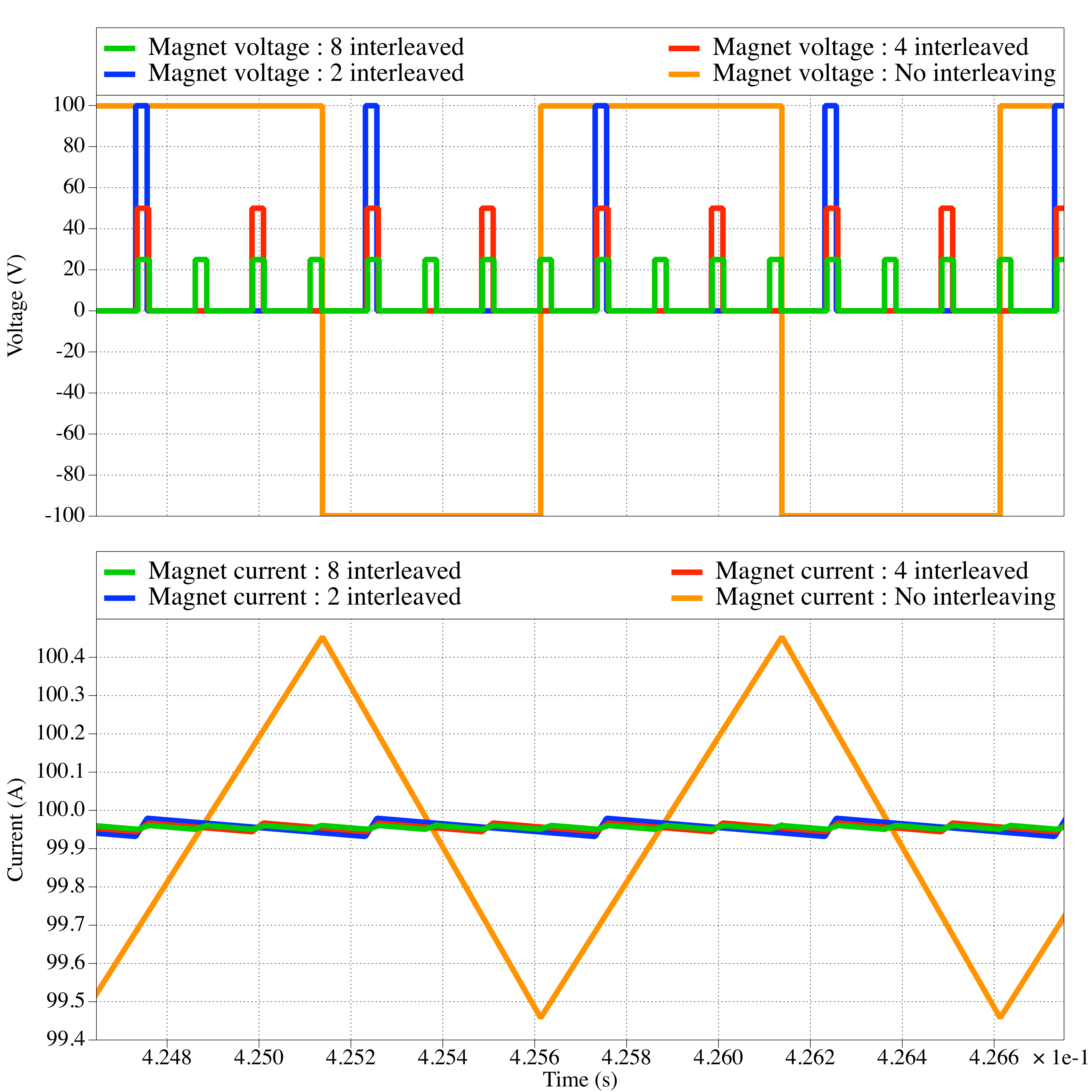}
\caption{\label{fig:sim_results} A plot showing the wave-forms of the interleaved voltage and current supplied to the magnet.}
\end{figure}

\section{Effect of parallelisation and frequency on the current ripple}
Using the voltage and current results shown in Fig. \ref{fig:sim_results}, it is possible to estimate the ripple of the current through the magnet. If the number of interleaved bricks is larger than one ($N_{bricks}>1$), the voltage switches between zero and a positive value $V_M$, which is different to the bus voltage $V_{bus}$. And as the number of parallel connections increases, the number of pulses increases proportionally and the magnitude of the voltage is inversely proportional. The current in the magnet is ramped up when the voltage is positive, and slowly discharges through the resistor when the voltage is zero. This voltage is determined by $V_M = \frac{V_{arm}}{N_{arms}}$, and the duration of the applied voltage is controlled by the duty cycle of the bricks. The current is in the flat-top phase, so the average current over one cycle is equal to the DC-current applied, where the duty cycle $m$ gives the following average voltage: $\hat{V}_M = V_M m$. This is the voltage driving the ramp-up of the current. The voltage  $\hat{V}_M$ is also equal to the average DC-voltage required to sustain the current in the resistor component of the magnet, thus  $\hat{V}_M=I_{flat}R_M$.

Using the fundamental relationship of an inductor in \eqref{eq:Ind}, and considering the delta current during the increasing of the ripple, the increasing current $di_M = \Delta I_{ripple}$ can be expressed in terms of the voltage $V_M$, inductance $L_M$ and time $dt = m \frac{1}{f_{sw}}$.

\begin{equation}
    \label{eq:Ind}
    V_M = L_M \frac{di_M}{dt}
\end{equation}

\begin{figure}[htbp]
\centering 
\includegraphics[width=.6\textwidth]{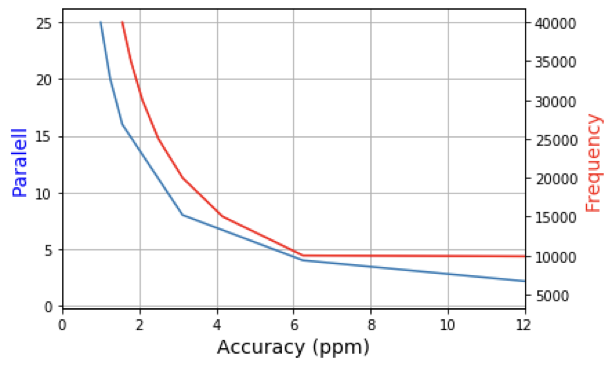}
\caption{\label{fig:results} Effect of interleaving on magnet voltage (and current) precision (blue line: varying number of bricks with a fixed frequency of 40kHz. Red line: varying the frequency with a fixed number of 16 interleaved bricks).}
\end{figure}

Using \eqref{eq:Ind} and expressing for the $\Delta I_{ripple}$ the ripple can be expressed as in Eq. \eqref{eq:I_ripple}. This finding closely matches the observed results from the simulations as shown in Fig. \ref{fig:results} for the case where $N_{arms}>1$. Using the results obtained in Eq. \eqref{eq:I_ripple}, the estimate is that 4.44 parallel strains should give the required accuracy at 50kHz.This result has been achieved with only a moderate increase of the switching frequency that becomes possible with a SiC technology device. In practice, 5 parallel interleaved bricks are sufficient to achieve the ripple accuracy of 0.9ppm. This topology is shown in Fig. \ref{fig:topology}.

\begin{equation}
\label{eq:I_ripple}
    \Delta I_{ripple} = \frac{V_M dt}{L_M} = \frac{\frac{\hat{V}_M}{m N_{arms}} \frac{m}{f_{sw}}}{L_M} = \frac{I_{flat}R_M}{L_M f_{sw} N_{arms}} \\
    \Longrightarrow \Delta I_{ppm} \approx \frac{R_M}{L_M f_{sw} N_{arms}}
\end{equation}

\section{Summary}
\label{sec:summary}
The current ripple form a single converter can be approximated by \eqref{eq:I_ripple_short}, where the ripple is inversely proportional to the switching frequency of the DC-DC converter. By using SiC MOSFETs it becomes possible to increase the switching frequency from 5kHz that is currently used, up to 60kHz and beyond. This gives an increase in accuracy by an order of magnitude or more. In addition, higher switching frequency results in a higher frequency on the ripple, reducing the size of any filters if they are still required. Thus, the anticipated cost and losses of filters are reduced. It is also possible to consider 3-level modulation (unipolar switching), reducing the ripple even further.
\begin{equation}
\label{eq:I_ripple_short}
    \Delta I_{ppm} \approx \frac{R_M}{L_M f_{sw} N_{arms}}
\end{equation}
Introducing parallel connected devices leads to an increased investment cost for the devices, but it can reduce the device losses, cooling requirements, filter requirements and filter losses. In addition to being a more scalable topology, which enables it to cover a large range of loads with minimal overcapacity and the possibility to have redundancy in the converter. It also introduces more segmentation for the storage in the case of pulsed loads with storage requirements for the DC-DC converter. By taking a complete view and considering lifetime costs, SiC MOSFETs seem to be very suitable technology for this application \cite{paper1}. Using the topology in Fig. \ref{fig:topology} with interleaving, it is possible to achieve the sub ppm filter requirement without using 
active filters for this particular load. The results in Fig. \ref{fig:results} shows the effect of interleaving and switching frequency on the current ripple. At relative small number of parallel connected bricks, the sub-ppm accuracy seems achievable. Combining with SiC MOSFETs this solution compares well to a conventional converter with IGBTs and active filters.

\bibliographystyle{JHEP.bst}

\providecommand{\href}[2]{#2}\begingroup\raggedright\endgroup

\end{document}